\documentclass[12pt]{article}
\usepackage{latexsym,graphicx,amssymb,color,amsmath}
\usepackage{epsfig}
\date{}
\graphicspath{}
\DeclareGraphicsExtensions{.eps}
\begin{document}

\vspace*{-2cm}
\begin{flushright}
q-bio.BM/0510028\\
DCPT--05/47\\
\end{flushright}

\vspace{0.3cm}

\begin{center}
{\Large {\bf Classification of capped tubular viral particles  \\ \vspace{0.3cm}
in the family of Papovaviridae}}\\ 
\vspace{1cm} {\large \bf T.\ Keef\,\footnote{\noindent E-mail: 
{\tt tk506@york.ac.uk}}, A.\
Taormina\,\footnote{\noindent E-mail: {\tt anne.taormina@durham.ac.uk}} and
R.\ Twarock\,${}^{1,}$\footnote{\noindent E-mail: 
{\tt rt507@york.ac.uk}}}\\
\vspace{0.3cm} {${}^1$}\em Department of Mathematics \\ University of York\\
\vspace{0.3cm} {${}^3$} Department of Biology \\ University of York \\
York YO10 5DD, U.K.\\ \vspace{0.3cm} {${}^2$\em \it Department of Mathematical
Sciences\\ University of Durham\\ Durham DH1 3LE, U.K.}\\ 
\end{center}

\begin{abstract}
\noindent A vital constituent of a virus is its protein shell, called the viral capsid, that encapsulates and hence provides protection for the viral genome. Viral capsids are usually spherical, and for a significant number of viruses exhibit overall icosahedral symmetry. The corresponding surface lattices, that encode the locations of the capsid proteins and intersubunit bonds, can be modelled by Viral Tiling Theory. \\
It has been shown in vitro that under a variation of the experimental boundary conditions, such as  
the pH value and salt concentration, tubular particles may appear instead of, or in addition to, spherical ones. In order to develop models  that describe the simultaneous assembly of both spherical and tubular variants, and hence study the possibility of triggering tubular malformations as a means of interference with the replication mechanism, Viral Tiling Theory has to be extended to include 
tubular lattices with end caps. This is done here for the case of Papovaviridae, which play a distinguished role from the  viral structural point of view as they correspond to all pentamer lattices, i.e. lattices formed from clusters of five protein subunits throughout. These results pave the way for a generalisation of  
recently developed assembly models. 
\end{abstract}
\newpage
\section{Introduction}
Viruses are fascinating simple organic systems, consisting of a very compact genome and a protective protein shell or capsid, which hijack host cells ten times their size in animals and plants, and which have the potential to kill. Advances in virology and the design of anti-viral therapeutics rely strongly on an understanding of the viral replication cycle, and, in particular, of the structure of capsids as well as of the mechanisms which trigger their assembly and disassembly. 

The importance of mathematical models for the structure of viral capsids and their assembly has been recognized for decades and is demonstrated by the significance of the 40-year old Caspar-Klug theory of quasi-equivalence \cite{CasparKlug} for the classification of viral capsids and 
for the three-dimensional reconstructions of viral capsids from experimental data \cite{Fuller:1999}. However, the existing models do not account for important classes of viruses including families of cancer-causing viruses whose study is of prime importance for the health sector. 

A very promising approach to solve the classification puzzle and yield a brand new perspective on viral structure is Viral Tiling Theory (VTT).   It was recently developed by Twarock \cite{Twarock:2004Virus}, initially in an attempt to address some of the shortcomings of the Caspar-Klug theory regarding viruses in the family of Papovaviridae. The distinctive feature of the latter is the fact that the surface lattices of their large icosahedral viral capsids are composed of clusters of five proteins only, while the Caspar-Klug theory predicts the presence of pentamers and hexamers. Within the VTT approach, the protein subunits composing the capsids are located in the corners of the tiles that meet at global and local symmetry axes.  For viruses in the family of Papovaviridae such as SV40, the tiles are kites and rhombs,  and the global 5-fold symmetry axes are located at vertices where five kites meet while the local 5-fold symmetry axes are located at vertices where two kites and three rhombs meet, as illustrated in  Fig.2 of \cite{KTT}. This reproduces the observed arrangements of protein subunits as pentamers. Crucially, VTT exploits the concept of symmetry to the full, and its group theoretical roots lie in the classification of all local symmetry axes of icosahedral structures, which are determined  via the affinisation of the non-crystallographic Coxeter group $H_3$ using a method inspired by the projection formalism known from the theory of quasicrystals and Penrose tilings \cite{Patera:2002, Twarocklocal:2005}. 
It is well-suited to the description of the capsid structure of Papovaviridae while still reproducing the tessellations relevant to the description of the viruses in the Caspar-Klug classification. Its predictive power is significantly enhanced, in comparison with the Caspar-Klug theory, through its ability to locate the {\em bonds} between protein subunits, and not only the location of the protein subunits themselves. 

Interestingly, VTT is also appropriate for the description of tubular variants of spherical capsids, whether they are capped or not.
This confirms the suggestion by Caspar and Klug that there should exist structural similarities between the surface lattices of tubular and spherical particles as the former occur alongside spherical shells during self-assembly \cite{Wikoff:1999}. While spherical particles can package the viral genome and are hence infectious, capped and non-capped tubular variants appear not to be experimentally\footnote{A. Zlotnick, private communication.}. It is therefore therapeutically desirable to derive an integrated model for assembly of spherical and tubular structures and track down which factors favour non-infectious malformations.

A first step in this direction is to classify tubular structures using VTT. 
Tubular capsid-like particles can be thought of as rolled sheets of lattices formed from capsomeres,
i.e. from clusters of protein subunits. There exists an abundance of tubular structures with open ends and these have been classified for Papovaviridae \cite{Twarock:2005}. On the other hand, tubular structures with end caps are more rare, as the existence of those places constraints on the possible surface structures of the tubes. It is the purpose of this paper to classify capped tubular structures and enumerate all possibilities  for the family of Papovaviridae as an example. This is important because {\em in experiments}, capped rather than open tubes are usually observed \cite{Kiselev:1969}. The classification derived here is a crucial input for assembly models which consider a scenario where both spherical capsids and tubes occur as products of assembly. It hence paves 
the way for a generalisation of the assembly models in \cite{KTT, KMT}.\\

After a brief overview of the Caspar-Klug theory and the discussion of generic features of capped tubular viral particles in section \ref{two}, a summary of the Viral Tiling Theory as applied to
spherical capsids in the family of Papovaviridae is presented in section \ref{three}. It provides the necessary tools for the classification of capped tubular structures for Papovaviridae, which is  organised according to the structures of the end caps and given in section \ref{four}. In section \ref{five} we 
conclude by discussing how these results can be implemented in the framework of assembly models.

\section{Icosahedral viral particles: generic features}\label{two}
The capped tubular variants of a given type of icosahedral virus are modelled as tubes with half of an icosahedral capsid closing each end. Their systematic classification may therefore be organized according to the type of end cap and tubular lattice compatible with the symmetries and the tessellation  of the 
corresponding spherical capsid considered. The size of the latter is characterized by a $T$-number introduced by Caspar and Klug  \cite{CasparKlug}. In the first subsection, we briefly review the ingredients of their classification system necessary to our purpose. We then discuss possible end caps given an icosahedral protein shell with number $T$. 

\subsection{Caspar-Klug spherical structures}
Spherical viral capsids with icosahedral symmetry  possess six 5-fold, ten 3-fold and fifteen 2-fold global rotation axes. Small capsids are organised in clusters of protein subunits centered on the global symmetry axes of the icosahedral structure. But these symmetry axes are not sufficient to accommodate the protein subunits of larger spherical viruses, which occur in multiples of 60. The {\em quasi-equivalence} theory advocated by Caspar and Klug consists in organising the protein clusters around the global symmetry axes, but also around some {\em local} symmetry axes (6-fold) of the icosahedral capsid. The underlying rule for generating large icosahedral capsids is as follows: the building blocks are not the protein subunits themselves, but instead {\em pentamers} and {\em hexamers}, which are regular polygonal groupings of five and six individual capsid proteins, called  {\em asymmetric} subunits because they are not mapped onto each other by the symmetry group. To build a capsid, one places a pentamer on each vertex of the icosahedron's six 5-fold symmetry axes (this accounts for $5\times 6\times 2=60$ individual subunits), and fills the space between pentamers with hexamers of the same side length. The bigger the capsid, the more hexamers one can fit, but only in the proportion of 12 to $10(T-1)$, where $T$ is called the {\em triangulation number}. This terminology stems from the fact that large capsids may be obtained by sub-triangulation of each face of the isosahedron in $T$ smaller facets, producing a shell geometrically equivalent to truncated icosahedra with $12+10(T-1)$ faces. The total number of protein subunits per capsid is therefore $12 \times 5 + 10(T-1)\times 6 = 60T$. $T$ may be written as $T=h^2+k^2+hk$ where the two non-negative integers $h$ and $k$ encode how to get from one pentamer to its nearest neighbour on the surface lattice that represents the organisation of the capsid. 
Low $T$-numbers are therefore
$T=1,3,4,7,9,...$. According to the above classification system,  large icosahedral capsids can be thought of as  subdivisions of the basic $T=1$ capsid, which is an icosahedron, i.e. the Platonic solid with twenty equilateral triangular faces. 

As already pointed out in the introduction, not all spherical icosahedral capsids exhibit the pattern of pentamers and hexamers predicted by Caspar and Klug. A prototype of such non-generic structures are viruses in the family of Papovaviridae, whose capped tubular variants  may be 
classified using Viral Tiling Theory as described in section \ref{four}. The classification of tubular variants of generic spherical particles is straightforward and given in the following subsection.

\subsection{Capped tubular variants of Caspar-Klug spherical viruses}
In classifying possible end caps, we find it convenient to consider first a $T=1$ capsid, and use its dual representation - the dodecahedron - whose twelve pentagonal faces represent the Caspar-Klug pentamers.  There are as many types of end caps as there are inequivalent ways  of cutting icosahedral capsids into equal halves, given the following two constraints. First of all,  each half spherical capsid must contain one end of each of the six global 5-fold symmetry axes in order to ensure  curvature is equally distributed over the end caps. This requirement stems from the fact that the capsid protein subunits cluster in groups of five around these symmetry axes to form the twelve curvature-generating pentamers of icosahedral capsids in the Caspar-Klug theory. This is analogous 
to pentagons providing curvature in capped carbon nanotubes \cite{Nanotubes}. Second of all, the six pentamers  within  each cap must be arranged symmetrically so that the cap's boundary can fit a tube-generating sheet with two parallel sides. All higher Caspar-Klug $T$-number viral capsids will have pentamers in the centre of the dodecahedral faces of a spherical dodecahedron surrounded by hexamers and as such are simple extensions of the dodecahedron.

\begin{figure}[ht]
\begin{center}
\includegraphics[width=12cm,keepaspectratio]{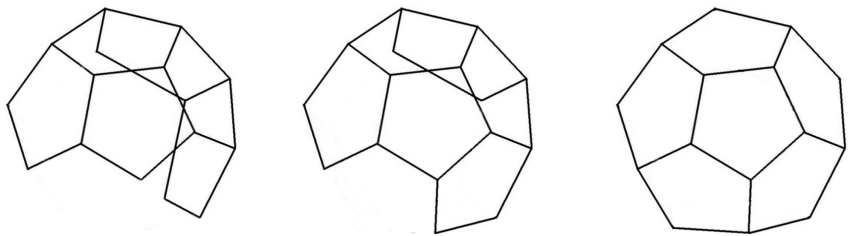}\\*
\end{center}
\caption{\em Half capsids cut about the 2-, 3-, and 5-fold axes.}
\label{fig1}
\end{figure}

These constraints yield precisely three types of end caps for each size of icosahedral capsid: they are obtained by cutting the capsid in half about the icosahedral 5-, 3-, and 2-fold vertices as shown in Fig.\ref{fig1}. 
It follows that capped tubular variants can be labelled by the $T$ number of the spherical particles associated with the end caps and the order $n=2,3$ or 5 of the symmetry axis about which the capsid is cut in half. For fixed $T$, there are therefore {\em at most} three types of end caps. Whether the three types occur in tubular variants depends on the compatibility of the end caps described above with
the structure of the tubular lattice, inherited from the structure of the associated icosahedral capsid. For instance, the second constraint imposed above ensures that the pentamers in the cap can be `glued' to the ends of {\em hexagonal} tubes, where the hexagons represent hexamers. Therefore, viruses within the classification of Caspar and Klug at number $T$ (whose capsids possess $10(T-1)$ hexamers around local 6-fold symmetry axes) admit capped tubular variants corresponding to the end caps labelled $(T,5), (T,3)$ and $(T,2)$. 

\section{Viral Tiling Theory and spherical particles in the family of Papovaviridae}\label{three}

Since a thorough understanding of some Papovaviridae viruses is extremely desirable in view of their cancer-causing nature, and since the Caspar-Klug theory fails to provide the necessary ingredients to grasp the structure of their capsids and to construct their tubular variants, we  
review here the elements of VTT required for the description of spherical particles in the family of Papovaviridae and, partially, for the discussion of their tubular malformations.
 
VTT provides a framework for the description of the surface lattices that act as 
blueprints for the organisation of the proteins in viral capsids. In particular, the surface lattices of the spherical particles are modelled as spherical tilings, i.e. tessellations in terms of a set of shapes (called tiles) that cover the sphere without gaps and overlaps. They are obtained via a group theoretical formalism that pinpoints the locations of 
{\em local} symmetry axes of order 3, 5 and 6, which mark the locations of the centers of the protein clusters (called trimers, pentamers and hexamers, respectively) that form the capsids. 

\begin{figure}[ht]
\label{figa}
\begin{center}
\raisebox{-1.3cm}{\includegraphics[width=5.1cm,keepaspectratio]{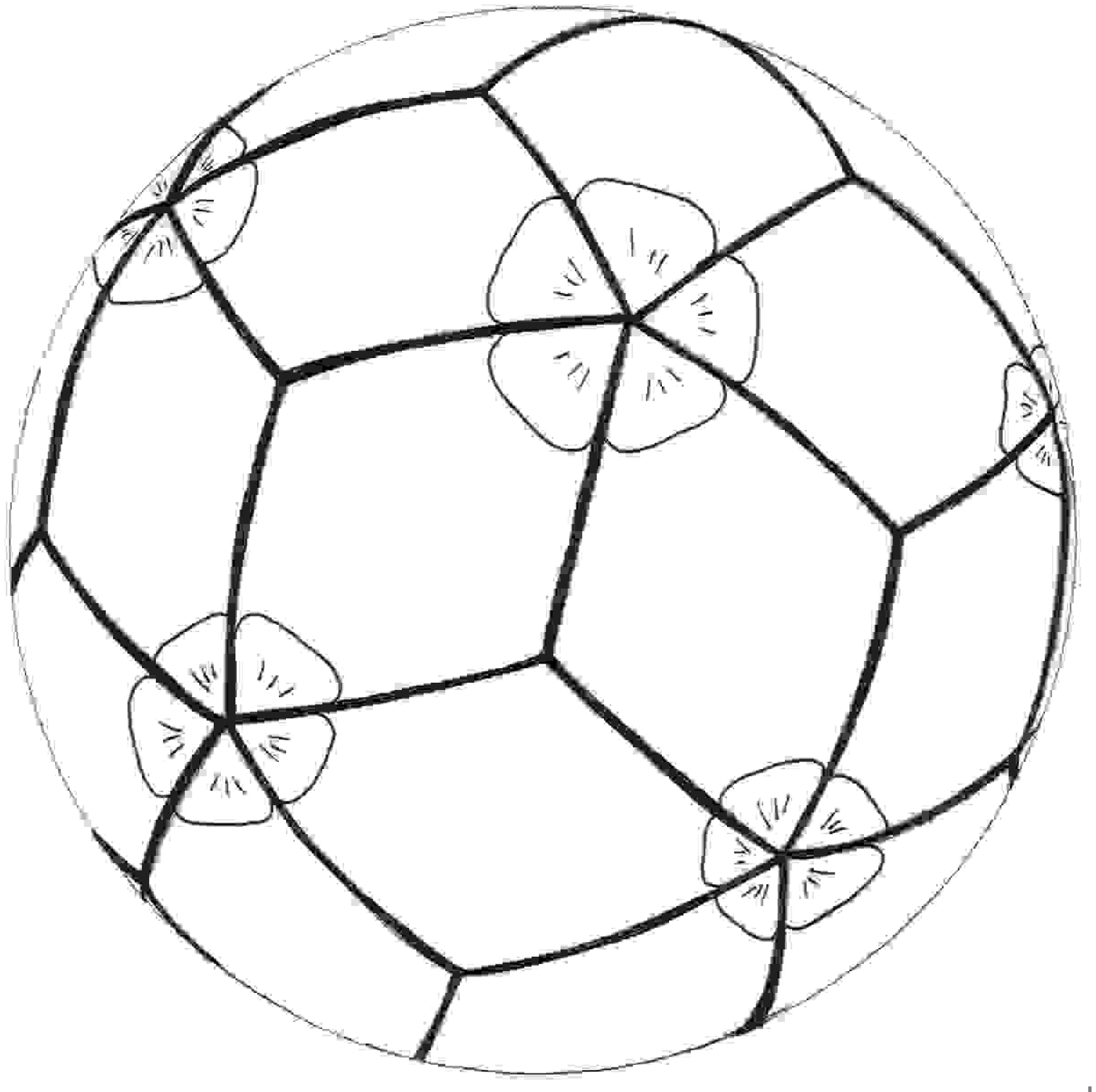}}\qquad
\includegraphics[width=4.5cm,keepaspectratio]{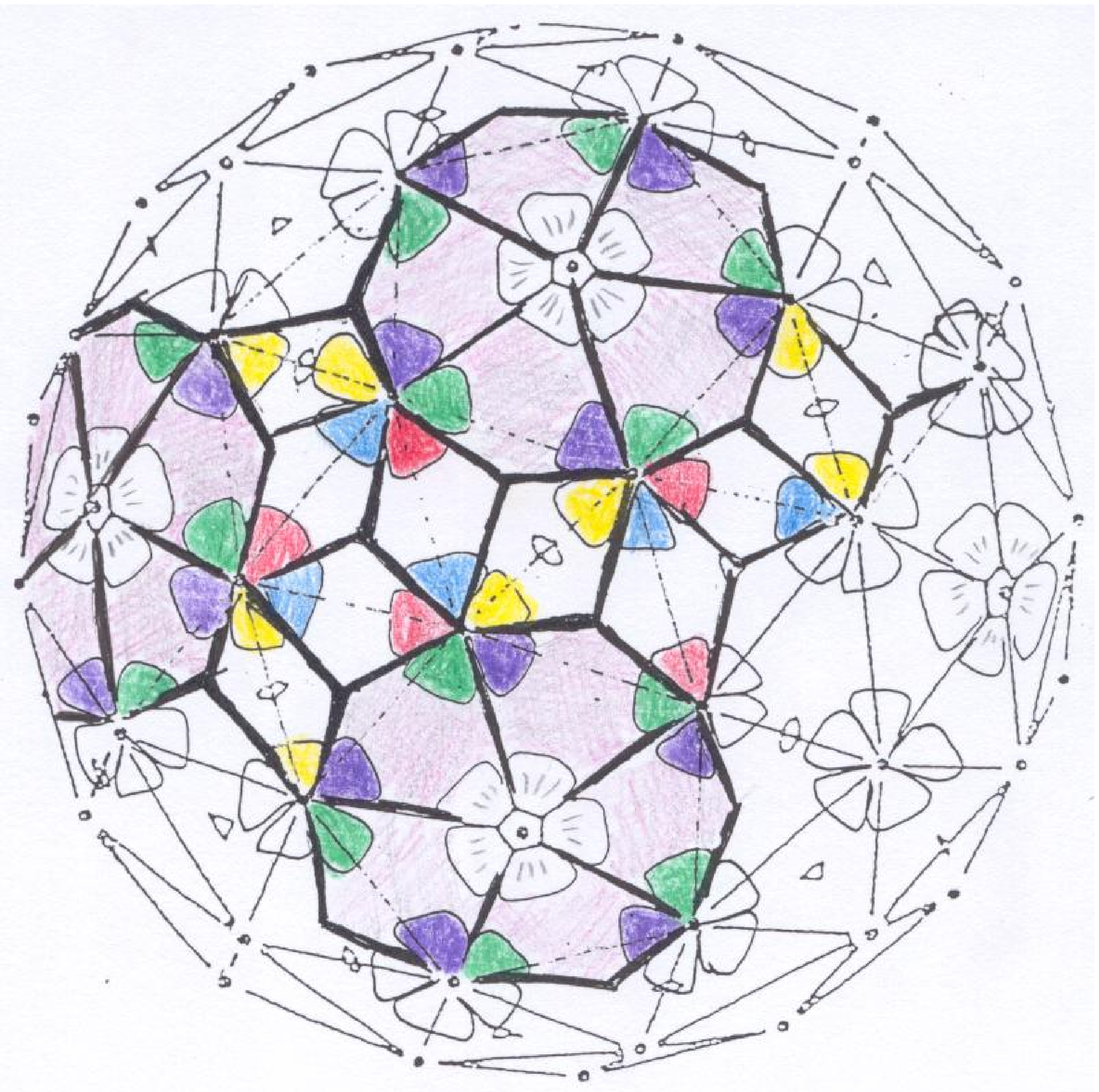}
\end{center}
\caption{The spherical tiling for a T=1 structure (left) and for a (pseudo-)T=7 particles (right) representing particles with icosahedral symmetry in the family of Papovaviridae.}
\label{spherical}
\end{figure}

A peculiarity of Papovaviridae lies in the fact that all protein clusters are pentamers, i.e. are arranged about local or global symmetry axes of order five. The icosahedral spherical particles in this family correspond to either configurations of 12 pentamers arranged around the symmetry axes of an icosahedron, or to configurations with 60 additional pentamers located around further local symmetry axes. The former are called $T=1$ structures according to the classification of Caspar and Klug, and the latter (pseudo-) $T=7$ structures because they share certain properties with $T=7$ geometries in their  classification. The corresponding tilings are shown in Fig. \ref{spherical}. The tiling corresponding to the (pseudo-)$ T=7$ particle has been derived in \cite{Twarock:2004Virus} and corresponds to a tessellation in terms of kites and rhombs. It describes the surface structure of viruses such as Polyomavirus, that has been a structural puzzle for a decade in view of Caspar-Klug theory (Liddington et al., 1991). 

Tiles can be interpreted from a biological point of view as representing interactions between the protein subunits located on the tiles. For example, in the case of the kite and rhomb tiling, kites correspond to the trimer interactions, that is interactions between three protein subunits, rhombs to dimer interactions, that is interactions between two protein subunits. The geodesics (arc) between the protein subunits on the tile in the spherical tiling mark the exact locations of the C-terminal arm extensions that form these bonds, as corroborated by the results in (Modis et al., 2002).

\section{Capped structures for pentagonal tubes: Papovaviridae variants}\label{four}

Pentagonal tubes are more rare than hexagonal ones. These are known to occur, for example, in the family of Papovaviridae, which we are discussing in this paper. A member of this family is Polyomavirus, which is a  (pseudo)-$T=7$ viral capsid whose capsid proteins form three capsid-like polymorphs as well as tubular structures \cite{Baker:1983}. The capsid-like polymorphs are a 72-capsomere (pseudo)-$T=7$ capsid, a 12-capsomere $T=1$ capsid and a 24-capsomere octahedral particle \cite{Salunke:1989}. Since the particles with octahedral symmetry are usually observed only in vitro, we are focussing on the icosahedral cases here. 

All capsomeres are pentamers built from five VP1 protein subunits. The surface structures of these polymorphs and all possible non-capped tubes may be described using VTT \cite{Twarock:2004Virus, Twarock:2005} as discussed in section \ref{three}. The relevant tiles are kites and rhombs, with protein subunits located at some specific vertices of each tile as shown in Fig.~\ref{fig2}: the kites represent trimer interactions and the rhombs represent dimer interactions between the individual subunits.  The tiles can meet according to two simple rules, which were originally based on biologic assumptions: 
\begin{figure}[ht]
\begin{center}
\includegraphics[width=5cm,keepaspectratio]{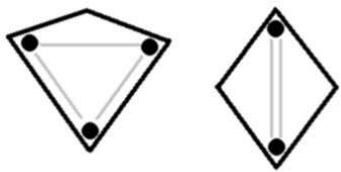}\\*
\end{center}
\caption{\em Tiles for models in the Papovaviridae family.}
\label{fig2}
\end{figure}
\begin{itemize}
\item {\bf \em Rule 1}: Edges may only meet if they are of the same length. This means that rhomb tiles may only meet kite tiles at the shorter edges of the kite tiles.
\item {\bf \em Rule 2}: If a vertex is decorated (has a protein subunit) then it may only meet another decorated vertex. Similarly, undecorated vertices may only meet other undecorated vertices. Partially decorated vertices cannot occur.
\end{itemize}
From a combinatorial point of view there are only a few possibilities for assembling tiles around a vertex according to these rules. These configurations are called { \em allowed vertex stars}. They are shown in Fig.~\ref{fig3}, bar the vertex stars which are only found in spherical caps. Consequently, any tiling obtained from kites and rhombs with the above matching rules has a periodic structure of ribbons of kites and rhombs.\footnote{Periodicity is a consequence of the matching rules; otherwise, aperiodic structures could also be constructed with this set of tiles.} A {\it tube of length $N$} is by definition a tube consisting of $N$ such ribbons.  Some examples of allowed tilings are shown in Fig.~\ref{fig4}. 
\begin{figure}[ht]
\begin{center}
\includegraphics[width=7cm,keepaspectratio]{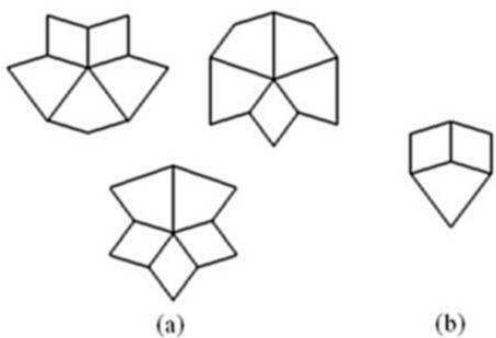}\\*
\end{center}
\caption{\em (a) The three allowed vertex stars with the vertex located on a pentamer. (b) The only other allowed vertex star.}
\label{fig3}
\end{figure}
\begin{figure}[ht]
\begin{center}
\includegraphics[width=7cm,keepaspectratio]{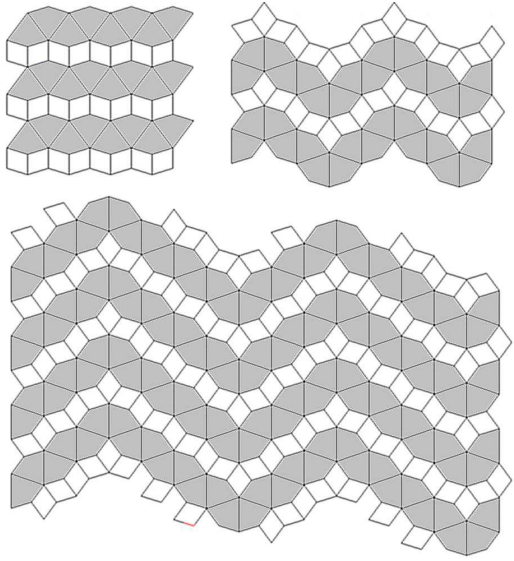}\\*
\end{center}
\caption{\em Three possibilities for lattices based on kites and rhombs, following the allowed tiling rules. Ribbons of kites have been highlighted to show the periodic structure.}
\label{fig4}
\end{figure}

In the following subsections we discuss which capped tubular structures can be associated with each spherical capsid in the family of Papovaviridae. We denote by $(T, n, N)$ the capped tubular variant associated with an icosahedral capsid of triangulation number $T$, whose end caps are 
obtained by cutting the capsid around a global $n$-fold symmetry axis and whose tubular section has
length $N$.

\subsection{T=1 Capped Tubes}
\begin{figure}[htb!]
\begin{center}
\includegraphics[width=16cm,keepaspectratio]{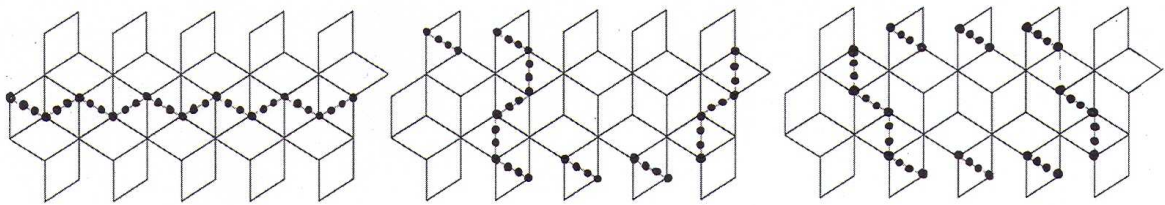}\\*
\end{center}
\caption{\em $T=1$ capsid-like particle made with rhombs only. The three ways of cutting the particle in half. From left to right: $(1,5)$ , $(1,3)$ and $(1,2)$.  This is a planar representation of a spherical capsid, so it can only be used as a qualitative diagram.}
\label{fig7}
\end{figure}

$T=1$ particles formed from the Polyomavirus capsid protein VP1 exhibit a structure with dimer interactions between protein subunits and therefore, according to VTT, may be tessellated by rhombs (see Fig.~\ref{spherical}). The surface lattice of such particles can be represented in planar geometry as shown in Fig.~\ref{fig7}, where we display the three ways of cutting the particles into halves according to the three options discussed in section \ref{two}. 

For each of them we must check whether the two halves can be completed to a tube by adding kite and rhomb tiles according to the allowed rules. 

\begin{figure}[htb!]
\begin{center}
\includegraphics[width=13cm,keepaspectratio]{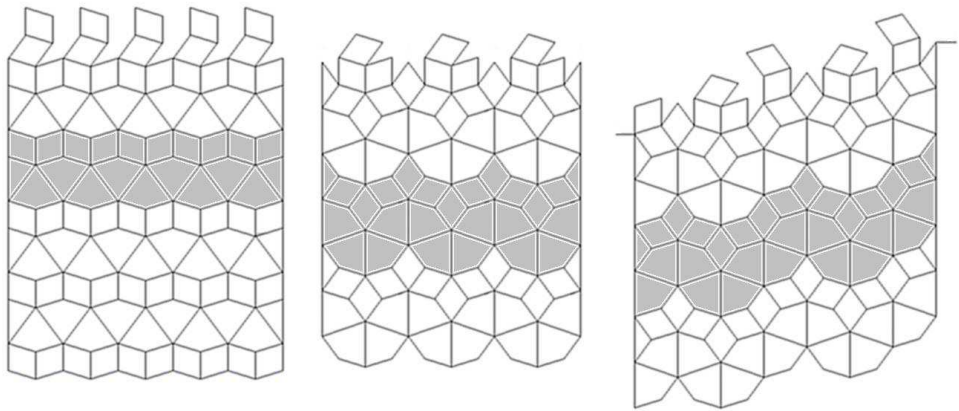}\\*
\end{center}
\caption{\em Nets for the three possible $T=1$ tubes. (a) $(1,5)$, (b) $(1,3)$, (c) $(1,2)$, where the stars show where to join the sides of the tube together. The highlighted bands show the repeated sections.}
\label{fig5}
\end{figure}

For $T=1$ particles, capped tubular variants can be constructed for any of the three types of end caps. Samples of the corresponding  tubes are shown in Fig.~\ref{fig5}. The $(1,5,N)$ capped tubular structure exhibits a total number of $12+10N$ pentamers as there are ten  in each highlighted ribbon and six in each end cap. The $(1,3,N)$ capped tubular structure contains  12 pentamers per ribbon and therefore a total of $12(N+1)$ pentamers, whereas the $ (1,2,N)$ capped tubular structure has fourteen pentamers per ribbon and a total of $12+14N$ pentamers. It follows that the number of protein subunits in each are $5(12+10N)$, $60(N+1)$ and $5(12+14N)$ respectively.
\subsection{(Pseudo-)T=7 Capped Tubes}

\begin{figure}[ht]
\begin{center}
\includegraphics[width=7cm,keepaspectratio]{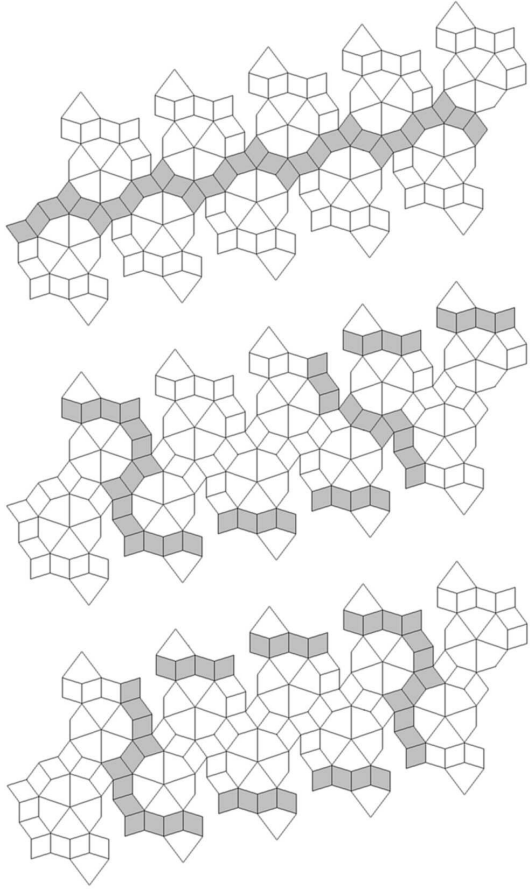}\\*
\end{center}
\caption{\em A $T=7$ lattice with rhombs highlighted to show the three possible ways of cutting it in half. The cuts run through the centres of the highlighted rhombs so that one half of each rhomb is contained in each cap. The configurations correspond to the $(7,5)$ cut (top), $(7,3)$ cut (middle) and $(7,2)$ cut (bottom) respectively.}
\label{fig8}
\end{figure}

The second type of spherical icosahedral particles in the family of Papovaviridae are the  (pseudo-)$T=7$ capsids in Fig.~\ref{spherical}, Section \ref{three}. We represent the corresponding tiling, again in planar geometry, in Fig.~\ref{fig8}. The three different ways of cutting the lattices in half according to the options discussed in section \ref{two} are indicated  by ribbons of highlighted rhombs, with cuts running through the centres of the rhombs. 

\begin{figure}[ht]
\begin{center}
\includegraphics[width=9cm,keepaspectratio]{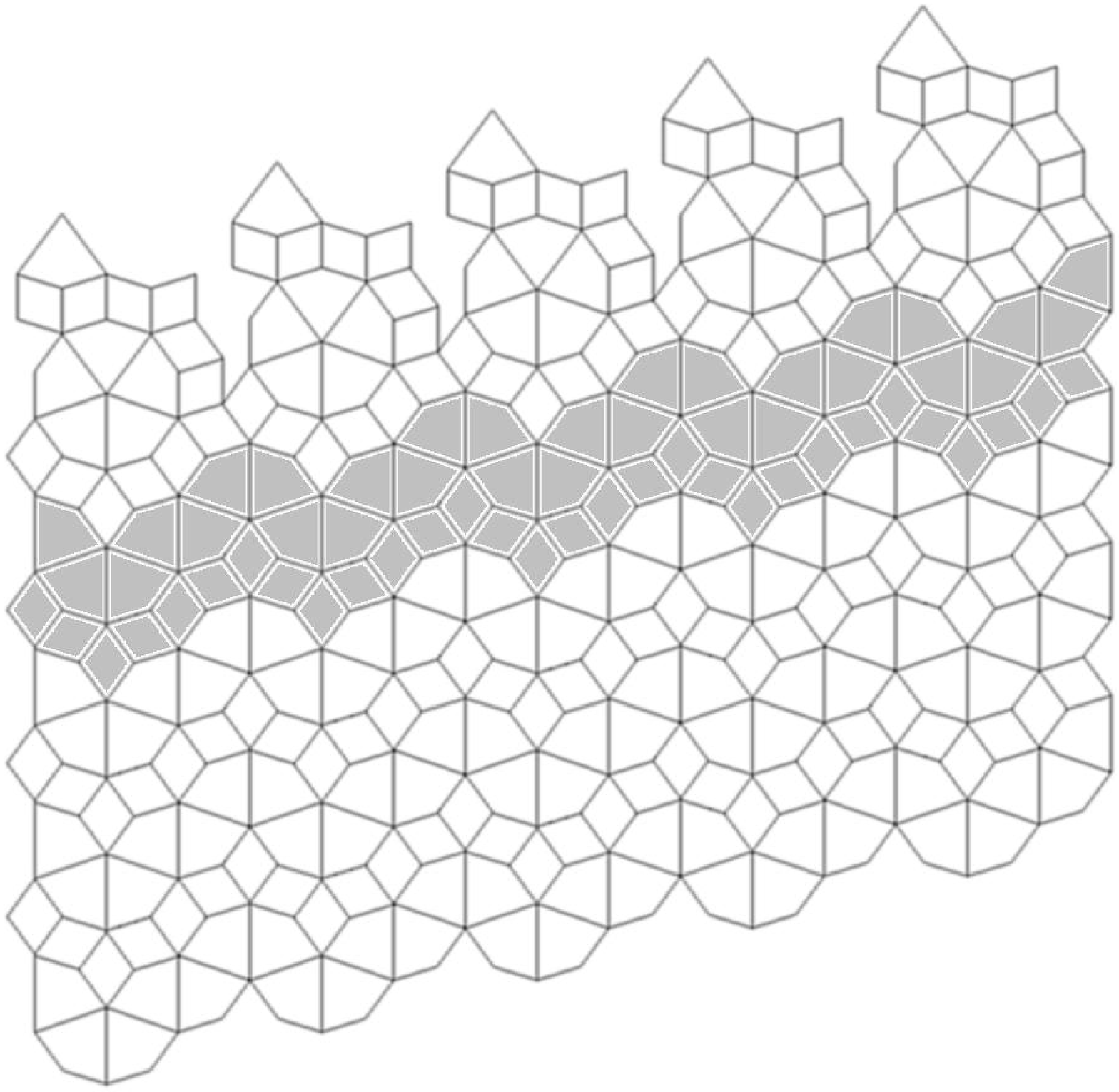}\\*
\end{center}
\caption{\em Net for the $(7,5)$ tube.}
\label{fig6}
\end{figure}

For each case one has to check whether these configurations can be continued into a tube using the vertex stars in Fig.~\ref{fig3}. This is possible for the $(7,5)$ cut, and the corresponding tube is shown in Fig.~\ref{fig6}, where periodically repeated sections have been highlighted. 
There are 30 pentamers per repeated section and thus, for a tube with $N$ repeated sections, there are $6(12+5N)$ pentamers in the complete shell.

\begin{figure}[ht]
\begin{center}
\includegraphics[width=8cm,keepaspectratio]{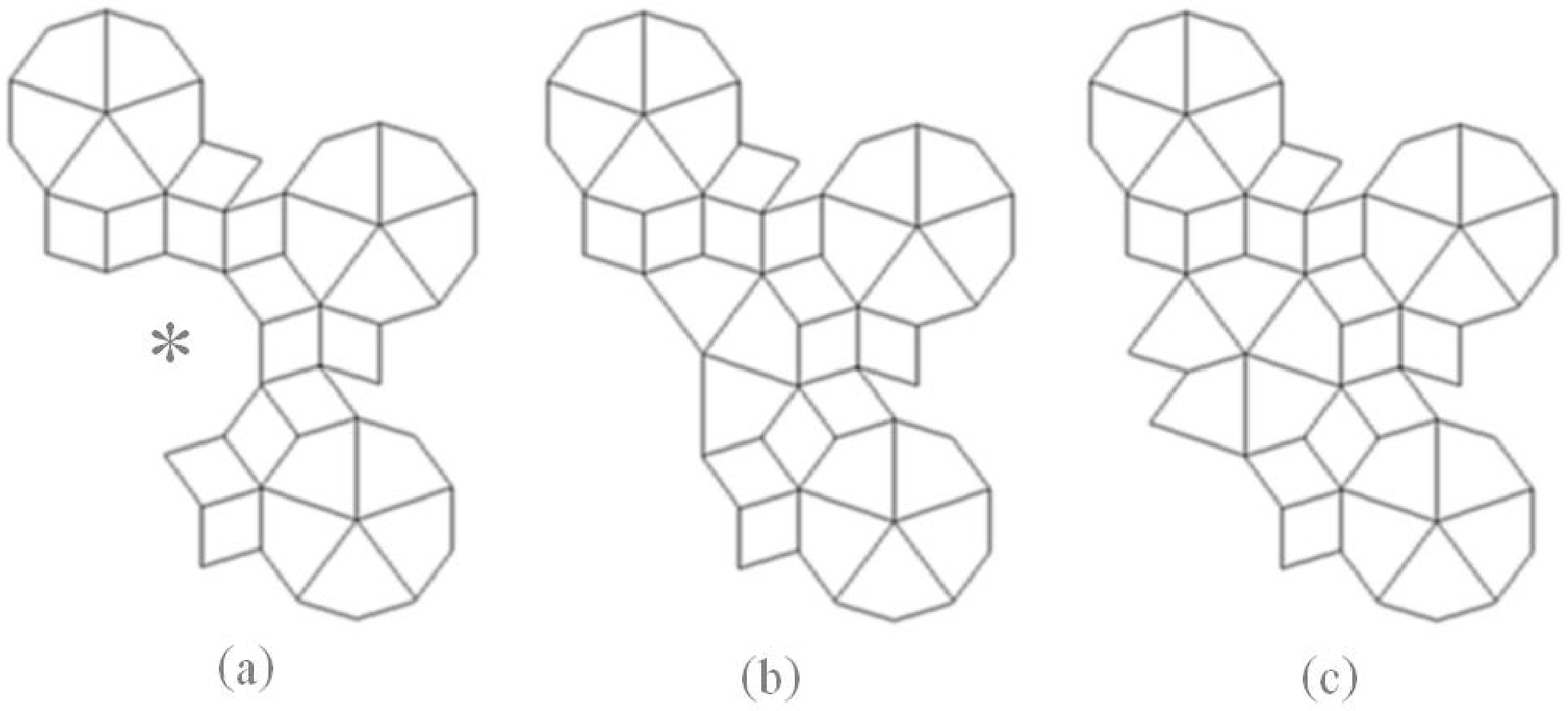}\\*
\end{center}
\caption{\em (a) $\star$ shows the untilable boundary section where the only choice is to add three kites as in (b). The only possible choice is to add two more kites. If all of the kites are pointing inwards, this configuration necessarily leads to a spherical capsid when continuing according to the rules. Otherwise, one obtains the untilable region shown in (c).}
\label{fig9}
\end{figure}

For the other two cuts, it is not possible to extend the end caps to tubes, because for both the $(7,3)$ cut and the $(7,2)$ cut there exists an area along the cut that corresponds to the scenario depicted in Fig.~\ref{fig9} on the left. 

The only possible way to complete this configuration based on Rule 1 above is to add three kite tiles as shown in Fig.~\ref{fig9}(b). According to Rule 1, this configuration has to be continued with two further kite tiles. If these are pointing inwards, any continuation of this configuration according to the rules necessarily leads to a spherical capsid. The alternative is to insert them pointing outwards as shown in Fig.~\ref{fig9}(c). However, according to Rule 2, this is an untilable configuration, i.e. it cannot be completed with the set of tiles according to the rules: the only geometrically possible tile for this gap would be a rhomb, but this leads to a configuration that does not correspond to any of the allowed vertex stars in Fig.\ref{fig3} and is hence excluded.

\section{Conclusion}\label{five}

In this paper, we have extended Viral Tiling Theory to tubular structures with end caps. In particular, we have classified all tubular particles in the family of Papovaviridae according to the structure of their end caps. Four different tubular lattices occur - three of them with $T=1$ end caps and one with a (pseudo-) $T=7$ end cap. 

Papovaviridae are distinguished from other families of viruses by the fact that the surface lattices of both spherical and tubular particles are formed from pentamers throughout, that is from protein clusters of five individual protein subunits. In this respect, Papovaviridae are different not only from the Caspar-Klug cases but also from fullerenes, i.e. cages formed from carbon atoms that have a structure similar to the lattices occurring in the Caspar-Klug theory. For fullerenes, it is known that tubular lattices with end caps occur as well as spherical lattices. They are called carbon nanotubes, and a classification of these may be found in \cite{Saito:1998}. Viruses 
described by the Caspar-Klug theory have associated tubular structures that have the same surface lattices as carbon nanotubes, and a classification of these structures can hence be inferred from the literature on carbon nanotubes. We have therefore focussed in this paper on Papovaviridae, which fall out of the scope of the Caspar-Klug classification.

While fullerenes also exist in spiral and toroidal conformations (see for example Chapter 7 in \cite{Saito:1998}), such structures have not been observed in viruses. A reason for this may be the fact that heptagonal rings are needed in addition to the pentagonal and hexagonal ones in order to create the negative curvature necessary to form spiral and toroidal shapes. Heptagonal ring structures can be formed from carbon atoms, but capsid proteins of viruses have never been observed in clusters of seven. Therefore, unless such clusters can be enforced by protein engineering, analogs to spiral and toroidal fullerene lattices are unlikely to exist in viruses. In any case, they are impossible in the framework of all-pentamer lattices, which correspond to the case of Papovaviridae. 

The classification of capped tubular structures for Papovaviridae derived here together with the classification of the spherical particles in this family obtained in earlier work exhaust all possible configurations that may be formed from the major capsid proteins of the viruses in this family. Therefore, these results can be used to construct assembly models that are exhaustive in the sense that they take the formation of all possible capsid structures into account. The derivation of corresponding assembly models as extensions of \cite{KTT, KMT} is planned. 
We shall use such models to investigate how realistic it is to trigger the formation of tubular (rather than spherical) structures and hence influence the viral replication cycle. Since Papovaviridae contain cancer-causing viruses this research may benefit the development of anti-viral drugs in the long-term.


\begin{thebibliography}{10}
\bibitem{CasparKlug}
D.L.D. Caspar and A. Klug, Cold Spring Harbor Symp. Quant. Biol. \textbf{27} 1 (1962).
\bibitem{Fuller:1999}
T.S.~Baker {\em et al}, Microbiology and Molecular Biology Reviews {\bf 63}, 862 (1999). 
\bibitem{Twarock:2004Virus}
R.~Twarock, J. Theor. Biol. {\bf 226}, 477 (2004); Bull. Math. Biol. \textbf{68} (2005). 
\bibitem{KTT}
T. Keef,  A. Taormina and R. Twarock, \emph{ Assembly models for Papovaviridae based on tiling theory}, q-bio.BM/0508031, Phys. Biol. 2 (2005).
 \bibitem{Patera:2002}
J. Patera and R. Twarock, J. Phys. A, Math. Gen. \textbf{35}
1551 (2002).
 \bibitem{Twarocklocal:2005}
R. Twarock, {\em A toolkit for the construction of icosahedral particles with local symmetry axes}, q-bio.BM/0508015. 
\bibitem{Wikoff:1999}
W.~R.~Wikoff and J.~E.~Johnson, Current Biology \textbf{9} R296 (1999); J.A.G.~Briggs {\em et al}, EMBO Journal {\bf 22}, 1707 (2003).  
\bibitem{Twarock:2005}
R.~Twarock, \emph{Mathematical models for tubular structures in the family of Papovaviridae}, Bull. Math. Biol. (2005), {\em in press}.
\bibitem{Kiselev:1969}
N.~A.~Kiselev and A.~Klug, J. Mol. Biol. \textbf{40} 155 (1969).
\bibitem{KMT}
T. Keef, C. Micheletti and R. Twarock, {\em Master equation approach to the assembly of viral capsids}, q-bio.BM/0508030.
\bibitem{Nanotubes}
R. Saito, {\em Physical properties of carbon nanotubes}, World Scientific (1998).
\bibitem{Baker:1983}
T. S. Baker, D. L. D. Caspar and W. T. Murakami, {\em Polyoma virus `hexamer' tubes consist of paired pentamers}, Nature {\bf 303}, 446 (1983).
\bibitem{Salunke:1989}
D. M. Salunke, D. L. D. Caspar and R. L. Garcea, Biophys. J. {\bf 56}, 887 (1989).
\bibitem{Saito:1998}
R. Saito, G. Dresselhaus, M.S. Dresselhaus, {\em Physical properties of carbon nanotubes}, Imperial College Press, London (1998). 
\end{thebibliography}
\end{document}